\documentstyle[12pt,psfig,paspconf]{article}
\begin{document}

\begin{center}
{\large\bf
Late-type stars members of young stellar kinematic groups
}

{\bf D. Montes$^1$, A. Latorre$^1$, M.J. Fern\'{a}ndez-Figueroa$^1$}
\end{center}

\noindent
{\em
$^1$ Departamento de Astrof\'{\i}sica,
Facultad de F\'{\i}sicas,
 Universidad Complutense de Madrid, E-28040 Madrid, Spain
(dmg@astrax.fis.ucm.es)\\
}

\vspace*{0.6cm}

To be published  in ASP Conf. Ser., 
{\em Stellar clusters and associations: convection, rotation, and dynamos} 
(Second "Three-Islands" Euroconference) (May 25 - 28, 1999, Mondello, Palermo, Sicily, Italy),
R. Pallavicini, G. Micela and S. Sciortino eds.

\vspace*{2.5cm}


\baselineskip=0.6truecm

\large

\hrule
\vspace{0.2cm}
\begin{center}
{\huge\bf Abstract}
\end{center}

We have compiled  a catalog of
late-type stars (F5-M) member of representative
young disk stellar kinematic groups: the
Local Association (Pleiades moving group, 20 - 150 Myr),
Ursa Mayor group (Sirius supercluster, 300 Myr),
 and  Hyades supercluster (600 Myr).
Other moving groups as IC 2391 supercluster (35 Myr) and
Castor Moving Group (200 Myr) have been also included.

Stars have been selected from previously established member of
stellar kinematic groups based in photometric and kinematic properties
as well as from candidates based in other criteria as their
level of chromospheric activity, rotation rate, lithium abundance.
Precise measurements of proper motions and parallaxes taken from
Hipparcos Catalogue, and published radial velocity measurements
are used to calculate Galactic space motions (U, V, W)
in order to determine the membership of the selected stars to the different
stellar kinematic groups.

In addition to kinematic properties we also give for each star
photometric, spectroscopic and physical properties as well as
information about activity indicators and Li abundance.
Some chromospherically active binaries results to be also members
of some of these stellar kinematic groups.

\vspace{0.5cm}
\hrule

\newpage
\hrule
\vspace{0.2cm}
\begin{center}
{\huge\bf Introduction}
\end{center}

Stellar kinematic groups, moving groups or superclusters, are kinematically coherent groups of stars
that should have the same basic properties as a genuine open cluster (except for spatial compactness).
The origin of these groups can be the evaporation of a open cluster of the remnants of a star formation region.
It has long been known that in the solar vicinity there are several kinematic groups of stars that share
the same space motions that well know open clusters.
The best documented groups are  the  Hyades supercluster (Eggen 1992b) associated with the Hyades cluster (600 Myr), 
the Ursa Mayor group (Sirius supercluster) (Eggen 1984a, 1992a, 1998b, Soderblom \& Mayor 1993a, b) associated with the UMa cluster 
of stars (300 Myr). 
A younger kinematic group called the Local Association or Pleiades moving group seems to consists of a reasonably
coherent kinematic stream of young stars with embedded clusters and associations such as the Pleiades, $\alpha$ Per, 
NGC 2516, IC 2602 and the Scorpious-Centaurus cluster (Eggen 1983, 1992c).
The age of the star of this association range from about 20 to 150 Myr.
Evidences have been found that X-ray and EUV selected active stars and lithium-rich stars
(Favata et al. 1993, 1995, 1998; Jeffries \& Jewell 1993; Mullis \& Bopp 1994; Jeffries 1995)
are member of this association.
Other two young moving groups are the IC 2391 supercluster (35-55 Myr) (Eggen 1991, 1995) and 
the Castor Moving Group (200 Myr) (Barrado y Navacu\'es 1998).

Recently, several studies using extended samples of star with known radial velocities and astrometric data
taken from Hipparcos (Chereul et al. 1998, 1999, Dehnen 1998, Asiain et al. 1999) 
not only confirm the existence of classical moving groups, but also detect finer structures that in several cases
can be related to kinematic properties of nearby open clusters or associations.
More complex structures characterized by several longer {\em branches} (Sirius, middle, and Pleiades branches)
running almost parallel to each other across the UV-plane  
have been found by Skuljan et al. (1999) in their study of the velocity distribution of star in the solar neighborhood.

A large fraction of the well known members to the different moving groups are early type star, however few studies 
are centered in late-type stars. Identification of significant numbers of the late-type population of these 
young moving groups would be extremely important to the study of the chromospheric and coronal activity and their age evolution.

In this contribution we compile a sample of late-type stars, previously established members
or possible new candidates to different young stellar kinematic groups (see Table 1).
We examine their kinematic properties (distribution of stars in the velocity space, UV amd WV planes), 
 using the more recent radial velocities and astrometric data available,
in order to determine their  membership to the different moving groups.

\vspace{0.5cm}
\hrule

\newpage

\begin{table*}
\caption[]{Young stellar kinematic groups
\label{tab:skg}}
\begin{flushleft}
\normalsize
\begin{tabular}{llcccccccccccccccccccccccc}
\noalign{\smallskip}
\hline
\noalign{\smallskip}
Name & Cluster/s & Age   & U, V, W     &  \\
     &           & (Myr) & (km s$^{-1}$) &      \\
\noalign{\smallskip}
\hline
\noalign{\smallskip}
Local Association   & Pleiades, $\alpha$ Per & 20 - 150 & -11.6, -20.7, -10.5 &   \\
{\scriptsize (Pleiades moving group)} &  &    &              &   \\
IC 2391             & IC 2391          & 35   &  -22.4,  -17.5,  -9.4  &   \\
                    &                  &      &                           &   \\
Castor Moving Group &                  & 200  &  -10.7,  -8.0,  -9.   &   \\
                    &                  &      &              &   \\
Ursa Mayor group    &                  & 300  &  14.7, 1.5, -10.0  &   \\
{\scriptsize (Sirius supercluster)} &  &      &              &   \\
Hyades supercluster & Hyades, Praesepe & 600  & -40  -16  -3 &   \\
                    &                  &      &              &   \\
\noalign{\smallskip}
\hline
\end{tabular}

\end{flushleft}
\end{table*}


\newpage
\hrule
\vspace{0.2cm}
\begin{center}
{\huge\bf Selection of the Sample}
\end{center}

Stars included in this work have been selected from previously established member of
stellar kinematic groups (see references given in Table 1) based in photometric and kinematic properties
as well as from candidates based in other criteria as their
level of chromospheric activity, rotation rate, lithium abundance.
We have selected star from different sources:

\begin{itemize}

\item The study of Agekyan \& Orlov (1984) which searched for kinematic groups in the solar neighborhood.

\item The study of ages of spotted late-type stars by Chugainov (1991).

\item X-ray and EUV selected active stars and lithium-rich stars
(Favata et al. 1993, 1995, 1998; Jeffries \& Jewell 1993; Tagliaferri et al. 1994, Mullis \& Bopp 1994; 
Jeffries 1995, Schschter et al. 1996).

\item Single rapidly rotating stars as AB Dor, PZ Tel, HD 197890, RE J1816+541, BD+22 4409 (LO Peg), HK Aqr,
V838 Cen, V343 Nor, LQ Hya,
previously assigned membership of the Local Association.

\item Chromospherically active late-type stars dwarfs in the solar neighborhood with studied kinematic properties
(Young et al. 1987; Upgren 1988; Soderblom 1990; Ambruster et al. 1998).

\item Flare stars with studied kinematic properties (Poveda et al. 1996).

\item Other chromospherically active single and binary stars (Strassmeier et al. 1993, Henry et al. 1995, 1996;
Soderblom et al. 1998)

\item The study of nearby young solar analogs by Gaidos (1998).

\end {itemize}

\vspace{0.5cm}
\hrule

\newpage
\hrule
\vspace{0.2cm}
\begin{center}
{\huge\bf Membership to the moving groups}
\end{center}

In order to determine the membership of this sample to the different stellar kinematic groups we have studied the
distribution of stars in the velocity space by calculating the {\sc Galactic space-velocity components} (U, V , W)
in a right-handed coordinated system (positive in the directions of the Galactic center, Galactic rotation, and the
North Galactic Pole, respectively).
The procedures in Johnson \& Soderblom (1987) were used to calculate U, V, W, and their associated errors.

- {\sc Parallaxes} and {\sc proper motions} are taken from 
Hipparcos Catalogue (ESA, 1997); 
PPM (Positions and Proper Motions) Catalogue (R\"{o}ser \& Bastian 1991; Bastian et al, 1993; R\"{o}ser et al, 1994);
ACT Reference Catalog (Urban et al. 1997); and 
TCR (Tycho Reference Catalogue) (Hog et al. 1998). 

- {\sc Radial velocities} are taken primarily from the compilation 
WEB (Wilson Evans Batten) Catalogue (Duflot et al. 1995),
the Catalogue of radial velocities of Nearby Stars (Tokovinin, 1992), 
 and from other references given in SIMBAD, and in the
 CNS3, Catalogue of Nearby Stars, Preliminary 3rd Version (Gliese \& Jahreiss 1991).

\vspace{0.5cm}
    
-- In Fig 1. we represent the (U, V) and (W, V) planes (Boettlinger Diagram) for our star sample.
The distribution of the stars in this figure shows concentrations around the (U, V, W) position corresponding 
to the five moving groups listed in Table 1.
In base of these concentrations we have classified the stars of our sample as member of one of these moving groups or as 
other young disk star if their classification is not clear but it is inside or near the boundaries (dashed line in Fig. 1)  
that determine the young disk population as defined by Eggen (1984b, 1989).
In Fig. 2 we plot the (U, V) for the Local Association, with some star identified.

\vspace{0.5cm}

-- In Tables 2 to 6
\footnote{Tables 2 to 6 available at {\tt http://www.ucm.es/info/Astrof/ltyskg.html}} 
 we list the candidate stars for each moving group.
We give the name, spectral type, coordinates (FK5 1950.0), radial velocity (V$_{r}$) and the error in km/s,
parallax ($\pi$) and the error in milli arc second (mas),
proper motions  $\mu$$_{\alpha}$ and  $\mu$$_{\delta}$ and their errors in mas per year (mas/yr), and the
U, V, W, calculated components  with their associated errors in km/s.
In the last column we mark with Y previously established members of
the stellar kinematic group and Y? possible new members in base of their position in the (U, V) plane.

\vspace{0.4cm}
\hrule

\newpage

\hrule
\vspace{0.5cm}


\vspace{0.5cm}
\hrule

\newpage


\begin{figure*}
{\psfig{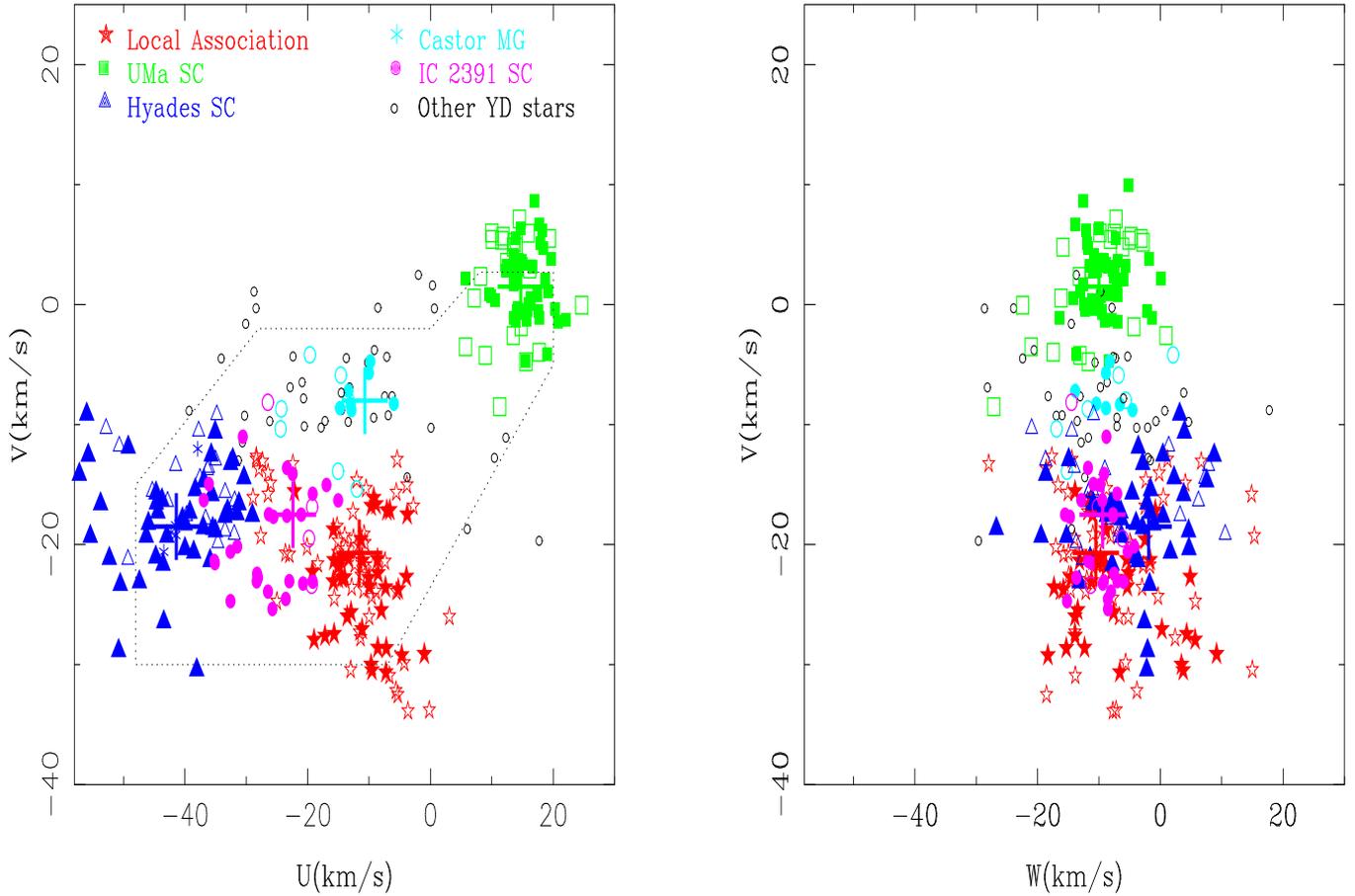}}
\caption[ ]{
(U, V) and (W, V) planes (Boettlinger Diagram) for our star sample.
We plot with different symbols and color the stars belonging to the different stellar kinematic groups,
and the other young disks stars.
Filled symbols are member stars (Y in tables) and open symbols are possible members (Y? in tables).
Big crosses are plotted in the central position of each group as given in Table 1.
The dashed line represent the  boundaries
that determine the young disk population as defined by Eggen (1984b, 1989).
}
\end{figure*}

\newpage

\begin{figure*}
{\psfig{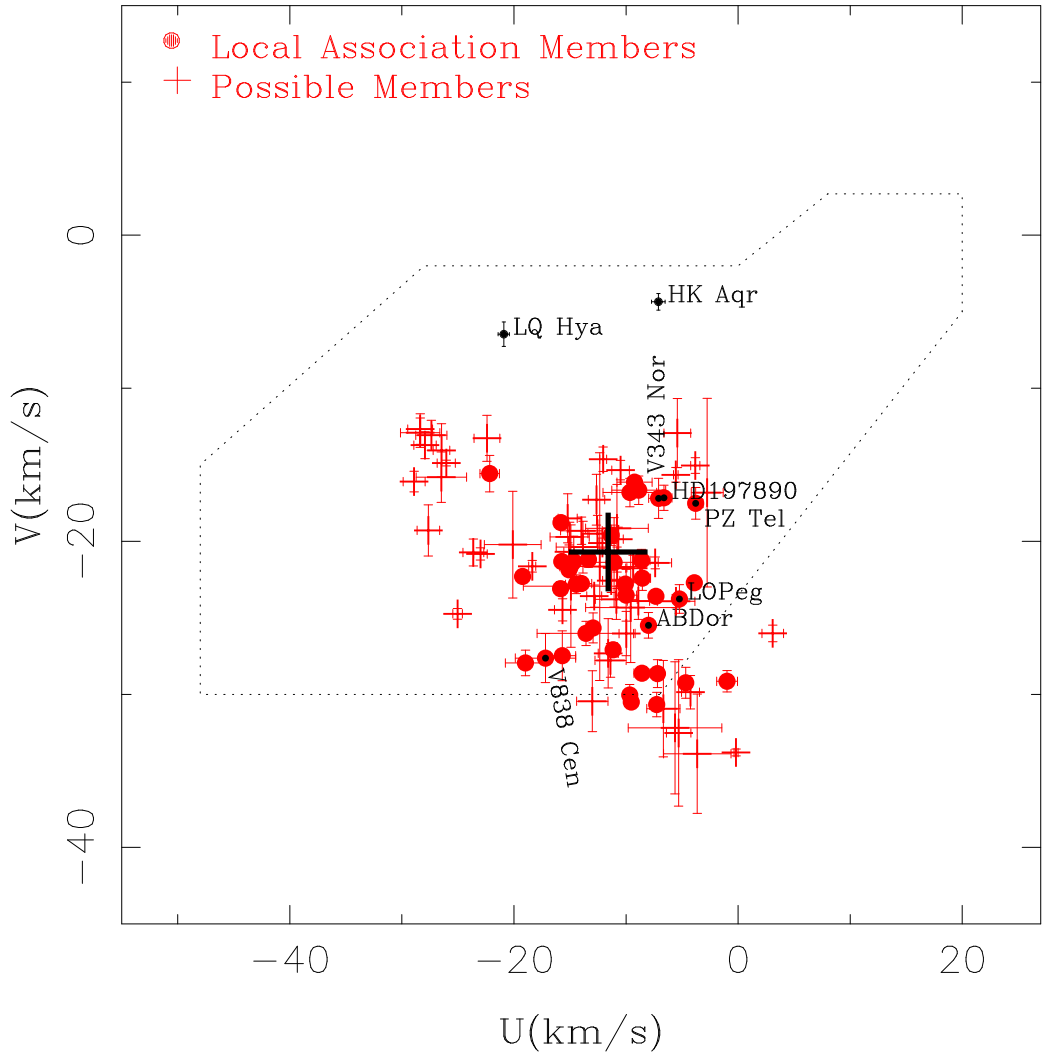}}
\caption[ ]{
(U, V) plane for Local association.
Filled symbols are member stars (Y in tables) and open symbols are possible members (Y? in tables).
A big cross is plotted in the central position of the group as given in Table 1.
The dashed line represent the  boundaries
that determine the young disk population as defined by Eggen (1984b, 1989).
We identified with their names the position of some previously established members.
The new calculated U and V values from LQ Hya and HK Aqr indicates that this stars are not members.
}
\end{figure*}

\end{document}